\documentclass[twocolumn,showpacs,preprintnumbers,
amsmath,amssymb,aps,prd,nofootinbib,superscriptaddress,
eqsecnum]{revtex4}
\usepackage{graphicx}

\makeatletter
\renewcommand{\p@subsection}{}
\makeatother

\newcommand{\Slash}[1]{\ooalign{\hfil/\hfil\crcr$#1$}}

\begin{document}

\title{The phase structure of a chiral model with dilatons in hot and dense matter}

\author{Chihiro Sasaki}
\affiliation{%
Frankfurt Institute for Advanced Studies,
D-60438 Frankfurt am Main,
Germany
}
\author{Igor Mishustin}
\affiliation{%
Frankfurt Institute for Advanced Studies,
D-60438 Frankfurt am Main,
Germany
}
\affiliation{%
Kurchatov Institute, Russian Research Center,
Moscow 123182, Russia
}

\date{\today}

\begin{abstract}
We explore the phase structure of a chiral model of constituent quarks
and gluons implementing scale 
symmetry breaking at finite temperature and chemical potential.
In this model the chiral dynamics is intimately linked to the trace anomaly 
saturated
by a dilaton field. The thermodynamics is governed by two condensates, 
thermal expectation values of sigma and dilaton fields, which are the order
parameters responsible for the phase transitions associated with the 
chiral and scale symmetries. Within the mean field approximation, 
we find that increasing temperature a system experiences a chiral phase 
transition and then
a first-order phase transition of partial scale symmetry restoration
characterized by a melting gluon-condensate takes place at a higher 
temperature.
There exists 
a region at finite chemical potential where the scale symmetry remains 
dynamically broken while the chiral symmetry is restored. 
%The model study
%suggests a gradual changeover of deconfined matter characterized by
%different melting temperatures of the composite operators. 
We also give
a brief discussion on the sigma-meson mass constrained from Lattice QCD.
\end{abstract}

\pacs{12.39.Fe,12.39.Mk,12.38.Mh}

\maketitle

%%%%%%%%%%%%%%%%%%%%%%%%%%%%%%%%%%%%%%%%%%%%%%%%%%%%%
\section{Introduction}
\label{sec:int}
%%%%%%%%%%%%%%%%%%%%%%%%%%%%%%%%%%%%%%%%%%%%%%%%%%%%

Effective theories of strongly interacting matter are
expected to capture non-perturbative aspects of QCD
in low-energy domain. 
They are constructed based on global symmetries of QCD Lagrangian
and their breaking pattern. In the limit of massless quarks 
the Lagrangian possesses the chiral symmetry and scale invariance, both
of which are dynamically broken in the physical vacuum due to the strong 
interaction. The QCD trace anomaly signals the emergence of a scale at the
quantum level from the theory without any dimension-full parameters~\cite{trace}.
Thus spontaneous chiral symmetry breaking, which gives rise to a nucleon
mass, and the trace anomaly are closely linked to each other~\cite{bardeen} and 
dynamical scales in hadronic systems are considered to originate from them.
How they behave under extreme conditions such as high temperature and density
is one of the main issues in QCD~\cite{review}.

The trace anomaly has been implemented in a chiral Lagrangian by introducing
a dilaton (or glueball) field representing the gluon condensate 
$\langle G_{\mu\nu}G^{\mu\nu} \rangle$~\cite{schechter}. Thermodynamics of the 
dilatons at
finite temperature and density has also been explored and the deconfinement 
phase transition was studied~\cite{campbell}. Incorporating the QCD scaling
properties into a non-linear chiral Lagrangian, the in-medium scaling associated
with chiral symmetry restoration, 
BR scaling~\cite{BR}, was introduced and some related works have been carried
out~\cite{kusaka,mishustin}. Besides, along with the Lattice QCD computations,
pure gluon dynamics at finite temperature has been formulated in several 
approaches~\cite{peshier,carter,qpm,miller,rg}.

In this paper we introduce a model of constituent quarks and gluons implementing
chiral and scale invariance in such a way that the model mimics the non-perturbative
nature of QCD in low energies. We will explore the thermodynamics and constrain
the sigma meson mass utilizing the QCD trace anomaly extracted from Lattice 
QCD~\cite{miller}. Imposing field theoretical 
requirements on the anomaly matching, we will give a suggestive phase diagram 
of QCD.

%%%%%%%%%%%%%%%%%%%%%%%%%%%%%%%%%%%%%%%%%%%%%%%%%%%%%
\section{A toy model}
\label{sec:model}
%%%%%%%%%%%%%%%%%%%%%%%%%%%%%%%%%%%%%%%%%%%%%%%%%%%%

In this section we briefly introduce our model for constituent quarks and gluons
restricting to a system with two flavors.

Scale invariance is implemented in a linear sigma model
via the following Lagrangian~\footnote{
 There are some uncertainties on introducing $\chi$ in the explicit breaking
 term. See e.g. \cite{gomm,BR}. This does not change our results. 
}:
\begin{eqnarray}
{\mathcal L}
&=&
\bar{q}i\Slash{\partial}q 
{}+ G_S\bar{q}\left( \sigma + i\vec{\tau}\cdot\vec{\pi} \right)q
\nonumber\\
&&
{}+ \frac{1}{2}\left( \partial_\mu\sigma\partial^\mu\sigma
{}+ \partial_\mu\pi\partial^\mu\pi \right)
\nonumber\\
&&
{}+ \frac{1}{2}\partial_\mu\chi\partial^\mu\chi
{}- V_\sigma - V_\chi\,,
\nonumber\\
V_\sigma
&=&
\frac{\lambda}{4}\left[ \left(\sigma^2 + \vec{\pi}^2 \right) 
{}- \sigma_0^2 \left(\frac{\chi}{\chi_0} \right)^2\right]^2
{}- \epsilon\left( \frac{\chi}{\chi_0}\right)^2\sigma\,,
\nonumber\\
V_\chi
&=&
\frac{1}{4}B\left( \frac{\chi}{\chi_0}\right)^4
\left[ \ln\left( \frac{\chi}{\chi_0}\right)^4 - 1 \right]\,,
\end{eqnarray}
where $G_S$ is the scalar coupling constant and $B$ is the bag constant.
All other notations follow the standard linear sigma model.
We assume that
the constituent gluons become massive due to the non-vanishing gluon
condensate, $\langle \chi \rangle \neq 0$. 
This is achieved by introducing the Lagrangian for the constituent gluon
field $A_\mu$,
\begin{equation}
{\mathcal L}_A
=
{}- \frac{1}{4}A_{\mu\nu}A^{\mu\nu}
{}+ \frac{1}{2}G_A^2\left(\frac{\chi}{\chi_0}\right)^2A_\mu A^\mu\,,
\end{equation}
with the field strength tensor 
$A_{\mu\nu} = \partial_\mu A_\nu - \partial_\nu A_\mu$ and
the coupling constant $G_A$ to the dilaton field.
The full Lagrangian is thus given by
\begin{equation}
{\mathcal L} \to {\mathcal L} + {\mathcal L}_A\,.
\end{equation}
Here we assume that the quarks have no direct coupling to the gluons
since
the interaction between the quarks and gauge fields is
embedded in $G_S$ and $G_A$.

Applying the mean field approximation, one finds the thermodynamic 
potential by performing the path integration over the quark and gluon
fields:
\begin{eqnarray}
\Omega
&=&
\Omega_q + \Omega_A + V_\sigma + V_\chi + \frac{1}{4}B\,,
\nonumber\\
\Omega_q
&=&
\gamma_q\int\frac{d^3p}{(2\pi)^3}T\left[ 
\ln\left( 1 - n_q \right) + \ln\left(1 - \bar{n}_q \right)
\right]\,,
\nonumber\\
\Omega_A
&=&
-\gamma_A\int\frac{d^3p}{(2\pi)^3}T
\ln\left(1 + n_A \right)\,,
\end{eqnarray}
with the degeneracy factors for quarks $\gamma_q = 2N_fN_c = 12$
and for gluons $\gamma_A = 2(N_c^2-1) = 16$. A constant term is added
so that $\Omega=0$ at $T=\mu=0$. The effective masses of
the quasi-particles are defined by
\begin{equation}
M_q = G_S\sigma\,,
\quad
M_A = G_A\frac{\chi}{\chi_0}\,.
\end{equation}
The thermal distribution functions are given by
\begin{eqnarray}
&&
n_q = \frac{1}{e^{(E_q - \mu)/T} + 1}\,,
\quad
\bar{n}_q = \frac{1}{e^{(E_q + \mu)/T} + 1}\,,
\nonumber\\
&&
n_A = \frac{1}{e^{E_A/T} - 1}\,,
\end{eqnarray}
with the quasi-particle energies
$E_q = \sqrt{|\vec{p}|^2 + M_q^2}$ and
$E_A = \sqrt{|\vec{p}|^2 + M_A^2}$.

The stationary condition, $\frac{\partial\Omega}{\partial\sigma} =
\frac{\partial\Omega}{\partial\chi} = 0$, leads to the following
coupled gap equations:
\begin{eqnarray}
&&
\gamma_q\int\frac{d^3p}{(2\pi)^3}
\frac{M_q}{E_q}G_S \left( n_q + \bar{n}_q \right)
{}+ \lambda\sigma\left[ \sigma^2 - \sigma_0^2\left(\frac{\chi}{\chi_0} 
\right)^2 \right]
\nonumber\\
&&
{}- \epsilon\left( \frac{\chi}{\chi_0}\right)^2 = 0\,,
\\
&&
\gamma_A\int\frac{d^3p}{(2\pi)^3}
\frac{M_A}{E_A}G_A n_A
{}-\lambda\sigma_0^2\left[ \sigma^2 - \sigma_0^2\left(\frac{\chi}{\chi_0}
\right)^2\right]\frac{\chi}{\chi_0}
\nonumber\\
&& 
{}- 2\epsilon\frac{\chi}{\chi_0}\sigma
{}+ B\left(\frac{\chi}{\chi_0}\right)^3
\ln\left(\frac{\chi}{\chi_0}\right)^4 = 0\,.
\end{eqnarray}

The mesonic parameters $\lambda$ and $\epsilon$ are related with the
sigma and pion masses and the pion decay constant via
\begin{eqnarray}
\lambda = \frac{m_\sigma^2 - m_\pi^2}{2f_\pi^2}\,,
\quad
\epsilon = m_\pi^2f_\pi\,,
\end{eqnarray}
where the vacuum sigma expectation value is $\sigma_0 = f_\pi$.
In the following calculation we will use $m_\pi = 138$ MeV and 
$f_\pi = 93$ MeV and alter the vacuum sigma mass $m_\sigma$ 
in the range $0.6$-$1.2$ GeV
because of its uncertainty.
The bag constant $B$ and dimensionful parameter $\chi_0$ are fixed
by the vacuum energy density ${\mathcal E} = \frac{1}{4}B = 0.76$
GeV fm$^{-3}$~\cite{narison} and the vacuum glueball mass $M_G = 1.7$ 
GeV~\cite{sexton} using the following definition:
\begin{equation}
M_G^2 = \frac{\partial^2V_\chi}{\partial\chi^2}
= \frac{4B}{\chi_0^2}\,.
\end{equation}
The coupling constants $G_S$ and $G_A$ are determined by requiring
that a nucleon is composed of three constituent quarks and a glueball
of two constituent gluons, thus,
\begin{eqnarray}
&&
M_q(T=\mu=0) = \frac{1}{3}m_N = 300\,\mbox{MeV}\,,
\nonumber\\
&&
M_A(T=\mu=0) = \frac{1}{2}M_G = 850\,\mbox{MeV}\,.
\end{eqnarray}

%%%%%%%%%%%%%%%%%%%%%%%%%%%%%%%%%%%%%%%%%%%%%%%%%%%%%
\section{Thermodynamics}
\label{sec:thermo}
%%%%%%%%%%%%%%%%%%%%%%%%%%%%%%%%%%%%%%%%%%%%%%%%%%%%

\begin{figure*}
\begin{center}
\includegraphics[width=5.5cm]{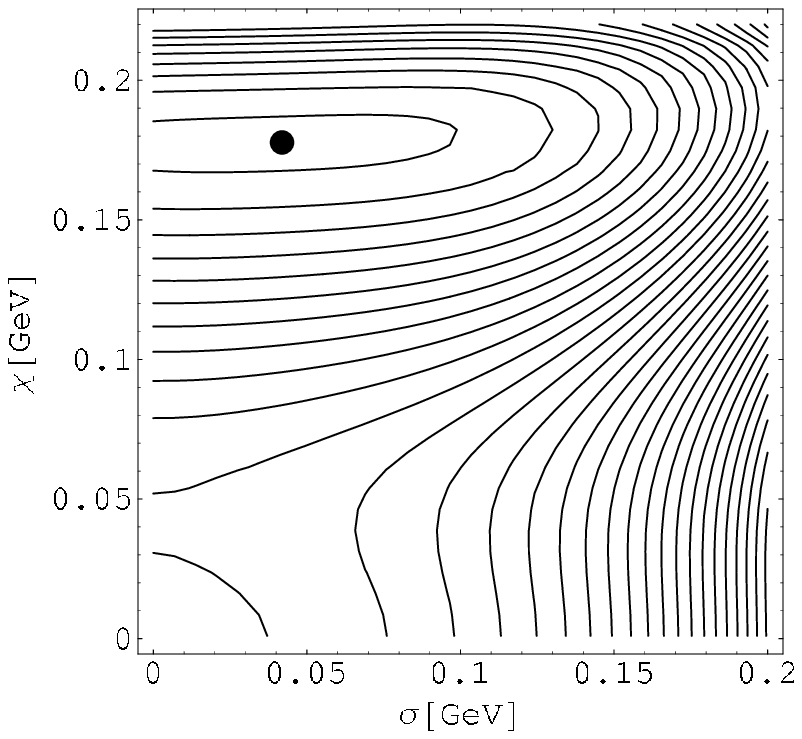}
\includegraphics[width=5.5cm]{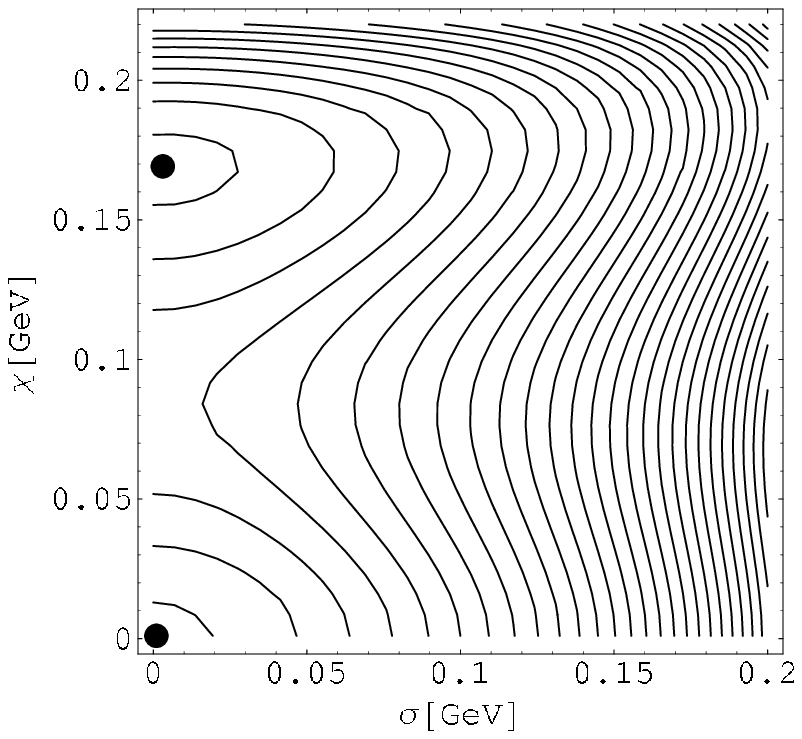}
\includegraphics[width=5.5cm]{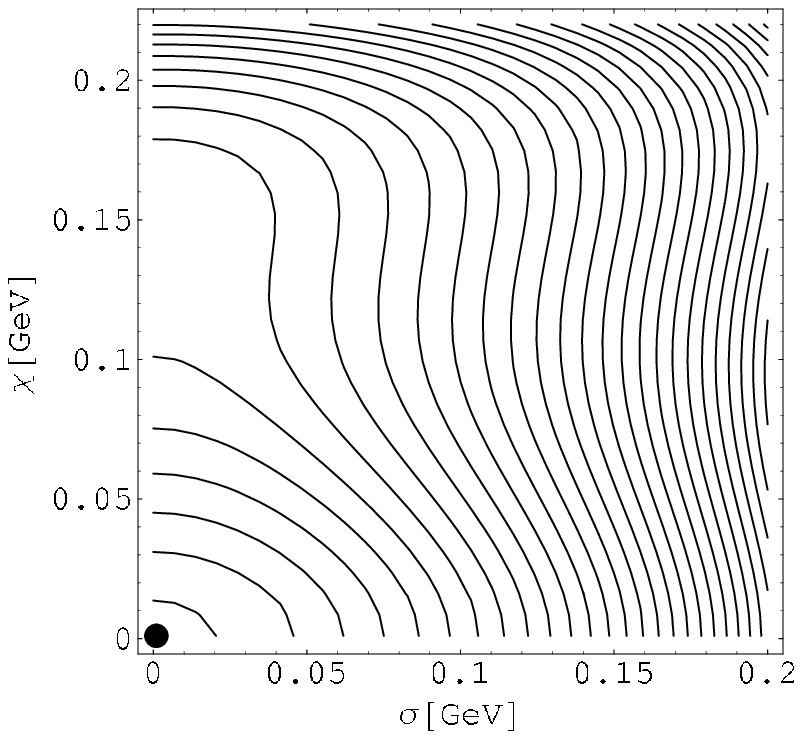}
\caption{
The contour plots of the thermodynamic potential at finite $T$ and $\mu=0$:
$T=153$ MeV (chiral crossover), $251$ MeV (first-order 
$\langle \chi \rangle \to 0$ transition)
and $300$ MeV from left to right. The black circle indicates the ground state.
$m_\sigma=600$ MeV at $T=0$ was used.
}
\label{contour}
\end{center}
\end{figure*}

The model introduced above describes the evolution 
of the two condensates, $\langle \sigma \rangle $ and $\langle \chi \rangle$, 
driven by temperature and chemical potential.
Figure~\ref{contour} shows the contours of the thermodynamic potential, 
taking the vacuum sigma mass being $m_\sigma = 600$ MeV
in $\sigma$-$\chi$ plane at $\mu=0$.
%%%%%%%%%%%%%%%%%%%%%%%%%%%%%%%%%%%%%%%%%%%%%
%%%%%%%%%%%%%%%%%%%%%%%%%%%%%%%%%%%%%%%%%%%
Increasing temperature from zero, first the system experiences partial
restoration of chiral symmetry at $T_{\rm chiral}$ indicated by the dropping 
$\sigma$ whereas another condensate $\chi$ remains almost a constant. Above
$T_{\rm chiral}$ the potential starts to exhibit a meta-stable state at
$\sigma \sim \chi \sim 0$ and a first-order phase transition takes place
at $T_{\chi=0}$ where the scale symmetry broken by non-vanishing $\chi$
is restored. Further above this temperature, the system remains at the 
trivial ground state.

\begin{figure}
\begin{center}
\includegraphics[width=7.5cm]{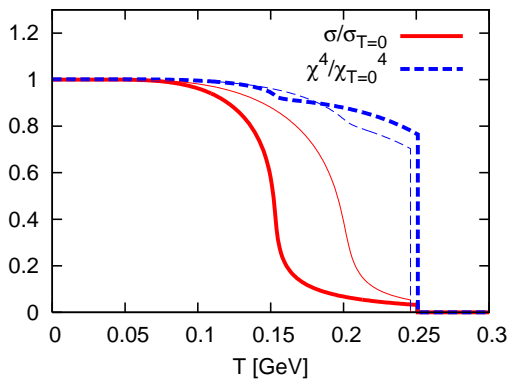}
\caption{
(Color online)
The normalized expectation values of $\sigma$ and $\chi^4$ fields
at $\mu=0$. The thick lines are calculated using $m_\sigma=600$ MeV at $T=0$
and the thin lines using $m_\sigma = 900$ MeV.
}
\label{cond}
\end{center}
\end{figure}

The thermal expectation values of $\sigma$ and $\chi$ obtained from the
gap equations in fact show a substantial reduction around the chiral
crossover and a jump at the first-order transition as seen in Fig.~\ref{cond}.
%%%%%%%%%%%%%%%%%%%%%%%%%%%%%%%%%%%%%%%%%%%%%
%%%%%%%%%%%%%%%%%%%%%%%%%%%%%%%%%%%%%%%%%%%
When the sigma meson is very massive, $\lambda \to \infty$, one finds
\begin{equation}
\langle \sigma \rangle
\simeq f_\pi\frac{\langle \chi \rangle}{\chi_0}\,,
\end{equation}
corresponding to non-linear realization of chiral Lagrangians, 
and the thermodynamics is governed
by a single condensate. Near the chiral symmetry restoration point the above
relation between the two condensates is not expected since the sigma meson cannot
be integrated out. The condensate of the dilaton field has a weak sensitivity 
to a temperature even above the chiral crossover and therefore it does not drive 
the disappearance of the chiral condensate. This feature however strongly depends 
on the sigma meson mass and for a larger $m_\sigma$ the gluon 
condensate is more affected by the chiral phase transition, as we will discuss
below.

\begin{figure}
\begin{center}
\includegraphics[width=7.5cm]{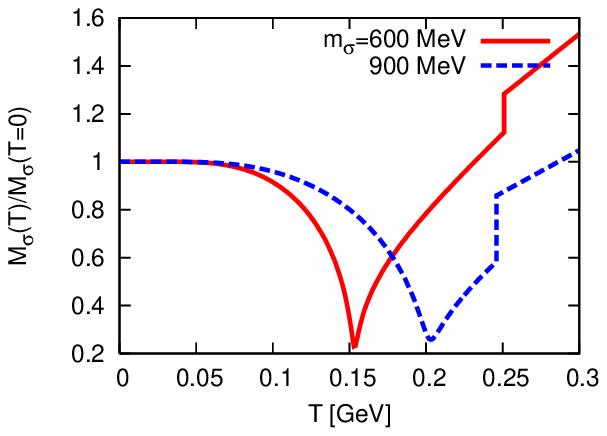}
\includegraphics[width=7.5cm]{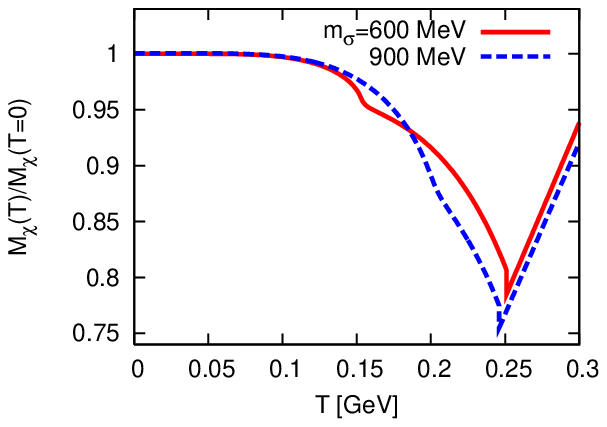}
\caption{
(Color online)
The thermal masses of $\sigma$ (top) and $\chi$ (bottom) fields at $\mu=0$.
}
\label{mass}
\end{center}
\end{figure}

In-medium masses of $\sigma$ and $\chi$ fields are defined by
\begin{equation}
M_\sigma^2 = \frac{\partial^2\Omega}{\partial\sigma^2}
\Big{|}_{\sigma=\langle\sigma\rangle\,, \chi=\langle\chi\rangle}\,,
\quad
M_\chi^2 = \frac{\partial^2\Omega}{\partial\chi^2}
\Big{|}_{\sigma=\langle\sigma\rangle\,, \chi=\langle\chi\rangle}\,.
\end{equation}
Their behavior as functions of temperature is given in Fig.~\ref{mass}.
%%%%%%%%%%%%%%%%%%%%%%%%%%%%%%%%%%%%%%%%%%%%%
%%%%%%%%%%%%%%%%%%%%%%%%%%%%%%%%%%%%%%%%%%%
Increasing temperature toward $T_{\rm chiral}$, $M_\sigma$ shows a strong
sensitivity to the phase transition as observed in the standard linear sigma 
models, whereas $M_\chi$ is rather modest. The two masses exhibit a jump when
$\chi$ vanishes. Above this temperature they follow a linear dependence of
temperature, $M_{\sigma,\chi} \sim T$, as expected.

\begin{figure}
\begin{center}
\includegraphics[width=7.5cm]{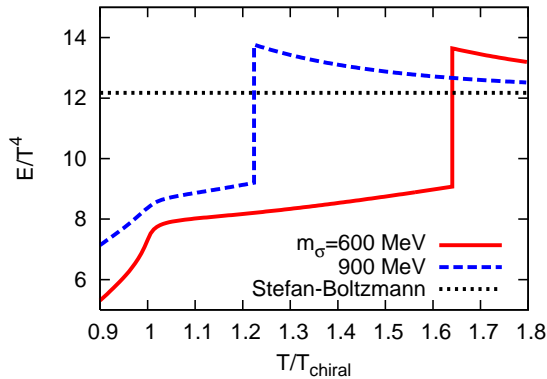}
\caption{
(Color online)
The scaled energy density at $\mu=0$.
}
\label{eos}
\end{center}
\end{figure}

In Fig.~\ref{eos} we show the energy density at $\mu=0$ as a function of temperature.
%%%%%%%%%%%%%%%%%%%%%%%%%%%%%%%%%%%%%%%%%%%%%
%%%%%%%%%%%%%%%%%%%%%%%%%%%%%%%%%%%%%%%%%%%
The standard linear sigma model (L$\sigma$M) almost follows the curves below 
$T_{\rm chiral}$, but strongly underestimates
the Stefan-Boltzmann 
(SB) limit that is a typical drawback of this model. Since the L$\sigma$M
Lagrangian does not contain gluons, bulk thermodynamics quantities are 
qualitatively in good agreement with the Lattice results when they are normalized 
by the SB limit for massless quarks, whereas not when normalized by the SB limit
for massless quarks and gluons. What we carried out in this paper is to improve 
the L$\sigma$M by introducing missing gluons. As shown in the figure the SB limit
is now reproduced. A defect to be removed is too strong
first-order phase transition even at $\mu=0$ which is absent in Lattice QCD.
Also, according to Lattice QCD the energy density should approach 
the SB limit from above.
We remark that direct comparison must be carried out in a more realistic 
framework beyond the mean field approximation. 
As shown in \cite{carter}, 
including thermal and quantum 
fluctuations of meson fields will be particularly important around $T_c$.

The trace anomaly exists at any temperature which is the only dimension-full 
quantity which breaks scale invariance of the theory explicitly. In our model, 
at high temperature and $M_q \ll T$ the pressure and energy density at 
$\mu=0$ are approximately expressed as
\begin{eqnarray}
P 
&=& 
\gamma_q\frac{7\pi^2}{720}T^4
{}- \frac{\gamma_q}{48}M_q^2T^2 - \frac{1}{4}B\,,
\nonumber\\
{\mathcal E}
&=&
\gamma_q\frac{21\pi^2}{720}T^4
{}- \frac{\gamma_q}{48}M_q^2T^2 + \frac{1}{4}B\,.
\end{eqnarray}
Consequently, one finds the trace anomaly (interaction measure) as
\begin{equation}
\Delta(T) = \frac{{\mathcal E} - 3P}{T^4} 
= \frac{B}{T^4} + \frac{\gamma_q M_q^2}{24T^2}\,.
\label{delta}
\end{equation}
%Introducing an effective interaction of quasi-particles developing with $T$, 
%$G_{S,A} \to G_{S,A}(T)$, which represents a residual interaction above the phase 
%transition point, will generate a further contribution and can match $\Delta$ 
%to the observation in lattice QCD~\cite{qpm} 
Lattice calculations~\cite{boyd} show that $\Delta$ has a non-perturbative term, 
$\Delta \sim 1/T^2$~\cite{pqm}. 
We see that this kind of contribution comes from the masses of quasiparticles.
However, the numerical value associated with the effective quark mass 
in Eq.~(\ref{delta}) is too small to explain this effect.
Fluctuations beyond the mean field approximation
will also contribute to the interaction measure~\cite{carter}.

\begin{figure*}
\begin{center}
\includegraphics[width=5.8cm]{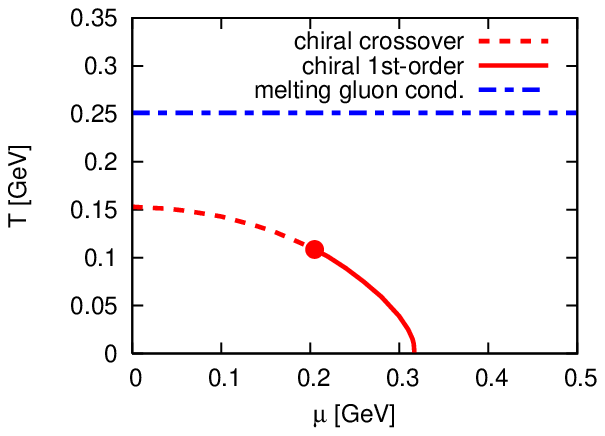}
\includegraphics[width=5.8cm]{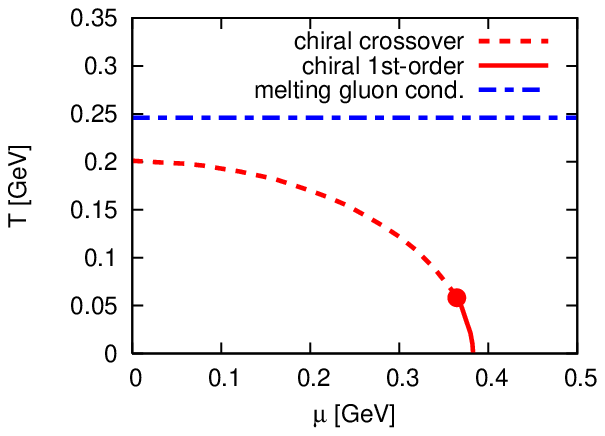}
\includegraphics[width=5.8cm]{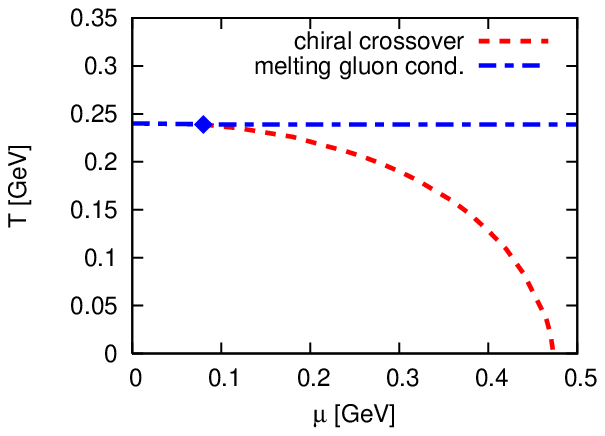}
\caption{
(Color online)
The phase diagram for different vacuum $m_\sigma$:
$m_\sigma = 0.6$ GeV (left), $0.9$ GeV (middle) and
$1.2$ GeV (right). The filled circle indicates the critical point
and the diamond the point where the first-order and crossover lines
intersect.
}
\label{phase}
\end{center}
\end{figure*}

Turning on the quark chemical potential $\mu$ practically does not affect 
the temperature at which the gluon condensate vanishes, $T_{\chi=0}$, 
whereas the chiral transition boundary exhibits an elliptic
shape and a critical point appears at an intermediate $\mu$, shown in Fig.~\ref{phase}
(left).
%%%%%%%%%%%%%%%%%%%%%%%%%%%%%%%%%%%%%%%%%%%%%
%%%%%%%%%%%%%%%%%%%%%%%%%%%%%%%%%%%%%%%%%%%
The boundary line of $T_{\chi=0}$ in general has a 
certain $\mu$ dependence via the gap equations. However, the sigma expectation 
value above $T_{\rm chiral}$ is small and little affects $\langle\chi\rangle$.
On the other hand, the chiral crossover line gets modified significantly
depending on $m_\sigma$ chosen in vacuum. For larger $m_\sigma$ the phase boundary
is systematically shifted to higher $T$ and $\mu$. The critical point also moves 
toward lower $T$ and eventually disappears from the phase diagram~\cite{lsmms}.
This is illustrated in Fig.~\ref{phase} (middle and right).
The thermodynamics at low temperature and high chemical potential is essentially
same as in the standard linear sigma model.

Making a matching of the trace anomaly between the model and QCD would
constrain a reliable range of $m_\sigma$. The divergence of the dilatation
current is given by~\cite{kusaka}
\begin{eqnarray}
\partial_\mu J^\mu
&=& - B\left(\frac{\langle\chi\rangle}{\chi_0}\right)^4
\nonumber\\
&&
{}+ \left( 4 - T\frac{\partial}{\partial T} 
{}- \chi\frac{\partial}{\partial\chi}\right) 
\Omega_A\big{|}_{\chi=\langle\chi\rangle}\,.
\label{traceD}
\end{eqnarray}
The left side of the above equation
is mostly saturated by the gluon condensate in QCD;
\begin{equation}
\partial_\mu J^\mu
= - \left( \frac{11}{24}N_c - \frac{1}{12}N_f\right)
\langle \frac{\alpha_s}{\pi} G_{\mu\nu}^a G^{\mu\nu}_a\rangle\,,
\label{traceG}
\end{equation}
where a small contribution due to the explicit breaking of chiral
symmetry is neglected. Lattice QCD calculations show that the thermal gluon condensate
decreases toward the pseudo-critical temperature of chiral symmetry
restoration and drops down to a half of its vacuum value at $T_{\rm chiral}$,
whereas it is quite stable at lower temperatures~\cite{miller}. This is also
a compatible feature with the QCD trace anomaly in terms of the soft and hard 
dilatons~\cite{miransky}, i.e. the disappearance of the soft dilaton is 
associated with chiral symmetry restoration and yields the melting gluon
condensate, or partial restoration of the scale symmetry breaking~\cite{LR}.
The two equations (\ref{traceD}) and (\ref{traceG}) tend to match
for a large $m_\sigma\sim 1$ GeV. With a small $m_\sigma$ the gluon condensate
does not show a significant drop at $T_{\rm chiral}$.
Thus, a rather heavy sigma-meson in the vacuum
seems to be favored by QCD, and this is a conceivable
scenario known from the vacuum phenomenology of the scalar mesons.
It should be noted that the matching is somewhat incomplete; Eq.~(\ref{traceD})
exceeds Eq.~(\ref{traceG}) by $\sim 15$\%. This may indicate that a stronger 
interaction between the quark and gluon sectors should be introduced. 
Besides, updating the
gluon condensate at finite temperature in Lattice QCD is necessary.

%%%%%%%%%%%%%%%%%%%%%%%%%%%%%%%%%%%%%%%%%%%%%%%%%%%%%
\section{Limit of infinitely heavy sigma meson}
\label{sec:heavy}
%%%%%%%%%%%%%%%%%%%%%%%%%%%%%%%%%%%%%%%%%%%%%%%%%%%%

\begin{figure}
\begin{center}
\includegraphics[width=7.5cm]{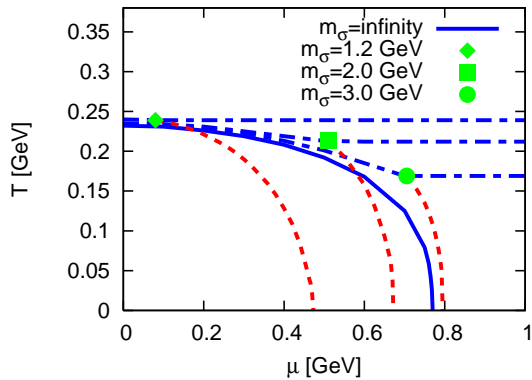}
\caption{
(Color online)
The phase diagram for different $m_\sigma$.
The line notation is same as in Fig.~\ref{phase}.
}
\label{phasems}
\end{center}
\end{figure}

It is instructive to study the phase diagram in the $\lambda \to \infty$
limit where the sigma meson becomes infinitely heavy. As discussed in the
previous section, the two critical temperatures, $T_{\chi=0}$ and $T_{\rm chiral}$,
get closer with increasing $m_\sigma$. With $m_\sigma \sim 1$ GeV they are almost
on top of each other and larger $m_\sigma$ yields an intersection of the first-order
phase transition of scale symmetry and chiral crossover lines at finite $\mu$.
This intersection moves to higher $\mu$ and lower $T$ for larger $m_\sigma$
as shown in Fig.~\ref{phasems}.
%%%%%%%%%%%%%%%%%%%%%%%%%%%%%%%%%%%%%%%%%%%%%
%%%%%%%%%%%%%%%%%%%%%%%%%%%%%%%%%%%%%%%%%%%
The boundary line of scale symmetry restoration is less sensitive to $\mu$
when the chiral symmetry is restored. This is because the major $\mu$-dependence
comes in via the sigma expectation value $\langle \sigma \rangle$ which is well
suppressed in restored phase. When $m_\sigma$ reaches infinity, the intersection
is kicked out and a single line of the first-order phase transition is left.
The region where chiral symmetry is restored whereas $\langle\chi\rangle \neq 0$
is unfavored in this limit.

The parameters of effective Lagrangians can alter with $T$ and $\mu$ since they
are obtained by integrating higher frequency modes out and thus expected to
carry information on the underlying QCD. Consequently, the phase diagram 
calculated with the parameters fixed using the vacuum quantities would be 
deformed and the first-order phase transition could remain on the phase diagram 
at high $\mu$ in a cold system.

%%%%%%%%%%%%%%%%%%%%%%%%%%%%%%%%%%%%%%%%%%%%%%%%%%%%%
\section{Implications for the QCD Phase Diagram}
\label{sec:phase}
%%%%%%%%%%%%%%%%%%%%%%%%%%%%%%%%%%%%%%%%%%%%%%%%%%%%

The present toy model exhibits three regions characterized by the two condensates:
(i) broken phase of chiral and scale symmetries, (ii) chirally restored but 
broken phase of the scale symmetry because of the non-vanishing 
$\langle \chi \rangle$, and (iii) chirally restored but explicitly broken phase 
of the scale symmetry by temperature.
What does the thermodynamics of the model suggest concerning the QCD phase
structure? Vanishing the condensate of the dilaton field indicates a disappearance
of the gluon composite at high temperature and its dissociation may signal a
transition of the system from the confined to deconfined phase. Thus, one
identifies the temperature $T_{\chi=0}$ with a temperature at which gluons are
released:
\begin{equation}
T_{\chi=0} \sim T_{\rm deconf}^{(g)}\,.
\end{equation}

\begin{figure}
\begin{center}
\includegraphics[width=7.5cm]{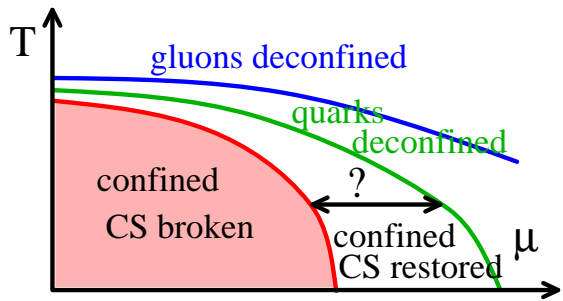}
\caption{
(Color online)
A sketch of the QCD phase diagram.
}
\label{phase2}
\end{center}
\end{figure}

The model yields a chiral transition temperature that is below $T_{\chi=0}$
in a wide range of the parameters. In $N_f=2$ QCD this is compatible with the 
anomaly matching which is often used to constrain possible massless excitations 
in quantum field theories~\cite{thooft}, and therefore the chirally restored 
phase with confinement is allowed. This suggests that the chiral symmetry
restoration takes place either below or at the deconfinement temperature, i.e.
\begin{equation}
T_{\rm chiral} \lesssim T_{\rm deconf}^{(q)}\,,
\end{equation}
where at $T_{\rm deconf}^{(q)}$ the quarks are released whereas the gluons remain
confined and it is not necessarily equal to $T_{\rm deconf}^{(g)}$.
As we have seen in the previous section, a large $m_\sigma$ can match with
the QCD requirement at $\mu=0$. This leads to the three distinct temperatures
which may be close to each other on the phase diagram. 
We note that this is consistent with the recent observation using a renormalization
group analysis where the fixed point of four-fermion interactions associated with
confinement plays an essential role~\cite{Braun}.
At finite $\mu$ no reliable constraint from QCD is known.
A suggestive phase diagram is given in Fig.~\ref{phase2}.
%%%%%%%%%%%%%%%%%%%%%%%%%%%%%%%%%%%%%%%%%%%%%
%%%%%%%%%%%%%%%%%%%%%%%%%%%%%%%%%%%%%%%%%%%

%%%%%%%%%%%%%%%%%%%%%%%%%%%%%%%%%%%%%%%%%%%%%%%%%%%%%%%
\section{Conclusions and Remarks}
\label{sec:conc}
%%%%%%%%%%%%%%%%%%%%%%%%%%%%%%%%%%%%%%%%%%%%%%%%%%%%%

In this paper we have studied thermodynamics and the phase structure of
a QCD-like model whose degrees of freedom are constituent quarks and gluons. 
Both chiral and scale symmetries are implemented in the model by introducing
mean fields representing $\bar{q}q$ and $G_{\mu\nu}G^{\mu\nu}$.
These symmetries are dynamically broken at low temperature and density.
The model thus mimics the features of QCD in the strong coupling region,
i.e. the spontaneous breaking of chiral symmetry and trace anomaly.
The results suggest that a system in deconfined phase develops gradually
with increasing temperature/density toward weakly-interacting quark-gluon 
matter composed of almost massless quarks and gluons.

The condensates of the sigma and dilaton fields are dynamically linked
via their gap equations. How strong they are correlated depends crucially
on the sigma-meson mass $m_\sigma$ chosen in vacuum. We found that a large 
$m_\sigma \sim 1$ GeV is consistent with the lattice result regarding 
the thermal behavior of the gluon condensate. 
This further
leads to the chiral phase transition which takes place almost simultaneously
with the deconfinement transition at $\mu \sim 0$. At finite $\mu$
these two transitions are expected to be separated.

In the scalar sector of low-mass hadrons, scalar quarkonium, tetra-quark 
states~\cite{jaffe} and glueballs are expected to be all mixed. 
How this can happen has been studied in certain simple models, see e.g.
~\cite{4QT} and references therein. It is an issue to be explored how the
presence of the tetra-quark modifies the phase structure presented in this work.

As an alternative approach one can use a parity doublet model
assuming a certain assignment of chirality to
nucleons with positive and negative parity~\cite{dk,mirror}.
As proposed in~\cite{dhls,dlfp},
the gluon condensate, 
more precisely the hard dilaton condensate, yields a chiral invariant mass of 
the nucleon, which stays 
non-vanishing above the chiral phase transition point. It is an interesting 
issue to explore the thermodynamics of a parity doublet model~\cite{pdm}
embedding dilatons and this will be reported elsewhere.

The present model can also be applied to a non-equilibrium system, where
the time evolution of the gluon condensate is described by the equation of
motion for the dilaton. On the other hand, in several models with Polyakov 
loops~\cite{pnjl,pqm,Megias} it is unclear how the kinetic term of the Polyakov 
loop dynamically emerges since the Polyakov loop by itself does not represent
a field but a character of the SU(3) color group. It would be interesting
to extend the work done in~\cite{Nahrgang} along this line.

%%%%%%%%%%%%%%%%%%%%%%%%%%%%%%%%%%%%%%%%%%%%%%%%%%%%%%%%
\subsection*{Acknowledgments}
%%%%%%%%%%%%%%%%%%%%%%%%%%%%%%%%%%%%%%%%%%%%%%%%%%%%%%

This work has been partly supported by the Hessian LOEWE initiative 
through the Helmholtz International Center for FAIR (HIC for FAIR),
and by the grants NS-7235.2010.2 and RFBR 09-02-91331 (Russia).

%%%%%%%%%%%%%%%%%%%%%%%%%%%%%%%%%%%%%%%%%%%%%%%%%%%%%%%%
%%%%%%%%%%%%%%%%%%%%%%%%%%%%%%%%%%%%%%%%%%%%%%%%%%%%%%%%%


\begin{thebibliography}{50}

\bibitem{trace}
  J.~C.~Collins, A.~Duncan, S.~D.~Joglekar,
  Phys.\ Rev.\  {\bf D16}, 438-449 (1977),
  N.~K.~Nielsen,
  Nucl.\ Phys.\  {\bf B120}, 212-220 (1977).

\bibitem{bardeen}
  W.~A.~Bardeen, C.~N.~Leung, S.~T.~Love,
  Phys.\ Rev.\ Lett.\  {\bf 56}, 1230 (1986).

\bibitem{review}
for recent reviews, see e.g.,
R.~S.~Hayano, T.~Hatsuda,
  Rev.\ Mod.\ Phys.\  {\bf 82}, 2949 (2010),
R.~Rapp, J.~Wambach, H.~van Hees,
  [arXiv:0901.3289 [hep-ph]],
W.~-G.~.Paeng, M.~.Rho,
  Mod.\ Phys.\ Lett.\  {\bf A25}, 399-422 (2010),
K.~Fukushima, T.~Hatsuda,
  Rept.\ Prog.\ Phys.\  {\bf 74}, 014001 (2011).

\bibitem{schechter}
  J.~Schechter,
  Phys.\ Rev.\  D {\bf 21}, 3393 (1980).

\bibitem{campbell}
  B.~A.~Campbell, J.~R.~Ellis, K.~A.~Olive,
  Nucl.\ Phys.\  {\bf B345}, 57-78 (1990);
  Phys.\ Lett.\  {\bf B235}, 325 (1990).

\bibitem{BR}
  G.~E.~Brown, M.~Rho,
  Phys.\ Rev.\ Lett.\  {\bf 66}, 2720-2723 (1991).

\bibitem{kusaka}
  K.~Kusaka, W.~Weise,
  Z.\ Phys.\  {\bf A343}, 229-234 (1992);
  Nucl.\ Phys.\  {\bf A580}, 383-407 (1994).

\bibitem{mishustin}
  I.~Mishustin, J.~Bondorf, M.~Rho,
  Nucl.\ Phys.\  {\bf A555}, 215-224 (1993).

\bibitem{carter}
  G.~W.~Carter, O.~Scavenius, I.~N.~Mishustin, P.~J.~Ellis,
  Phys.\ Rev.\  {\bf C61}, 045206 (2000).

\bibitem{peshier}
  A.~Peshier, B.~Kampfer, O.~P.~Pavlenko, G.~Soff,
  Phys.\ Rev.\  {\bf D54}, 2399-2402 (1996),
  P.~Levai, U.~W.~Heinz,
  Phys.\ Rev.\  {\bf C57}, 1879-1890 (1998).

\bibitem{qpm}
  A.~Dumitru, Y.~Guo, Y.~Hidaka, C.~P.~K.~Altes, R.~D.~Pisarski,
  Phys.\ Rev.\  {\bf D83}, 034022 (2011),
  P.~Castorina, D.~E.~Miller, H.~Satz,
  Eur.\ Phys.\ J.\  {\bf C71}, 1673 (2011).

\bibitem{miller} 
  D.~E.~Miller,
  Phys.\ Rept.\  {\bf 443}, 55 (2007).

\bibitem{rg}
B.~J.~Schaefer, O.~Bohr and J.~Wambach,
Phys.\ Rev.\ D {\bf 65}, 105008 (2002).

\bibitem{gomm}
  R.~Gomm, P.~Jain, R.~Johnson, J.~Schechter,
  Phys.\ Rev.\  {\bf D33}, 801 (1986).

\bibitem{narison}
  S.~Narison,
  Nucl.\ Phys.\ Proc.\ Suppl.\  {\bf 54A}, 238 (1997).

\bibitem{sexton}
  J.~Sexton, A.~Vaccarino and D.~Weingarten,
  Phys.\ Rev.\ Lett.\  {\bf 75}, 4563 (1995).

\bibitem{boyd}
  G.~Boyd, J.~Engels, F.~Karsch, E.~Laermann, C.~Legeland, M.~Lutgemeier, B.~Petersson,
  Nucl.\ Phys.\  {\bf B469}, 419-444 (1996).

\bibitem{lsmms}
B.~-J.~Schaefer and M.~Wagner,
Phys.\ Rev.\ D {\bf 79}, 014018 (2009).

\bibitem{miransky}
  V.~A.~Miransky and V.~P.~Gusynin,
  Prog.\ Theor.\ Phys.\  {\bf 81}, 426 (1989).

\bibitem{LR}
  H.~K.~Lee and M.~Rho,
  Nucl.\ Phys.\  A {\bf 829}, 76 (2009).

\bibitem{thooft}
G.~'t Hooft, in {\it Recent Developments in Gauge Theories},
ed. G.~'t Hooft {\it et al.} (Plenum Press, New York, 1980).

\bibitem{Braun} 
  J.~Braun and A.~Janot,
  arXiv:1102.4841 [hep-ph].

\bibitem{jaffe}
  R.~L.~Jaffe,
  Phys.\ Rev.\  {\bf D15}, 267 (1977);
  Phys.\ Rev.\  {\bf D15}, 281 (1977).

\bibitem{4QT}
  A.~Heinz, S.~Struber, F.~Giacosa and D.~H.~Rischke,
  Phys.\ Rev.\  D {\bf 79}, 037502 (2009).

\bibitem{dhls}
  C.~Sasaki, H.~K.~Lee, W.~G.~Paeng and M.~Rho,
  Phys.\ Rev.\  D {\bf 84}, 034011 (2011).

\bibitem{dlfp}
  W.~G.~Paeng, H.~K.~Lee, M.~Rho and C.~Sasaki,
  [arXiv:1109.5431 [hep-ph]].

\bibitem{dk}
  C.~E.~Detar and T.~Kunihiro,
  Phys.\ Rev.\  D {\bf 39}, 2805 (1989).

\bibitem{mirror}
  D.~Jido, M.~Oka and A.~Hosaka,
  Prog.\ Theor.\ Phys.\  {\bf 106}, 873 (2001).

\bibitem{pdm}
  T.~Hatsuda and M.~Prakash,
  Phys.\ Lett.\  B {\bf 224}, 11 (1989),
  D.~Zschiesche, L.~Tolos, J.~Schaffner-Bielich and R.~D.~Pisarski,
  Phys.\ Rev.\  C {\bf 75}, 055202 (2007),
  C.~Sasaki and I.~Mishustin,
  Phys.\ Rev.\  C {\bf 82}, 035204 (2010).

\bibitem{pnjl}
K.~Fukushima,
  Phys.\ Lett.\  B {\bf 591}, 277 (2004),
C.~Ratti, M.~A.~Thaler and W.~Weise,
  Phys.\ Rev.\  D {\bf 73}, 014019 (2006).

\bibitem{pqm}
  B.~-J.~Schaefer, J.~M.~Pawlowski, J.~Wambach,
  Phys.\ Rev.\  {\bf D76}, 074023 (2007).

\bibitem{Megias}
  E.~Megias, E.~Ruiz Arriola and L.~L.~Salcedo,
  Phys.\ Rev.\ D {\bf 74}, 065005 (2006).

\bibitem{Nahrgang}
  M.~Nahrgang, M.~Bleicher, S.~Leupold, I.~Mishustin,
  [arXiv:1105.1962 [nucl-th]].


\end{thebibliography}
\end{document}